# Object Kinetic Monte Carlo Simulations of Radiation Damage in Neutron-Irradiated Tungsten Part-I: Neutron Flux with a PKA Spectrum Corresponding to the High-flux Isotope Reactor

Giridhar Nandipati[a1], Wahyu Setyawan[a], Howard L. Heinisch[a], Kenneth J. Roche[a,b], Richard J. Kurtz[a] and Brian D. Wirth[c]

[a] *Pacific Northwest National Laboratory, Richland, Washington, USA*
[b] *Department of Physics, University of Washington, Seattle, WA 98195, USA*
[c] *University of Tennessee, Knoxville, Tennessee, USA*

**Abstract:** Object kinetic Monte Carlo simulations were performed to study the impact of varying dose rate (1.7 x$10^{-5}$–1.7 x $10^{-9}$ dpa/s) and grain size (2.0 and 4.0 μm in diameter) up to a dose of 1.0 dpa in pure, polycrystalline tungsten, subjected to a neutron irradiation having a PKA spectrum corresponding to the High Flux Isotope Reactor. The present study models defect cluster accumulation in tungsten, but does not consider the impact of transmutation or pre-existing defects beyond the grain boundary sinks, with varying grain size. With increasing dose rate, the vacancy cluster density increases, while the number density of vacancies decreases. Accordingly, the average vacancy cluster size and the fraction of vacancies that are part of visible clusters decreases with increasing dose rate. With increasing grain size, both the number densities of vacancies and vacancy clusters decrease, while both the fraction of vacancies in visible clusters and the average vacancy cluster size increase. This is caused by the pseudo-ripening of the vacancy clusters due to the longer-lived self-interstitial clusters in larger grains. The spatial ordering of vacancy clusters along {110} planes was observed for both grain sizes and all dose rates studied. Interplanar spacing increases with grain size; however, no clear dependence on dose or dose rate was observed. The results of this study show that 1D diffusion of self-interstitial clusters, while necessary, is not sufficient to form a void lattice, and that the diffusion of vacancies is also required. A methodology is suggested for choosing the simulation box dimensions so as to represent more faithfully the effects of one-dimensional migrating self-interstitial-atom clusters.

## 1. Introduction

Tungsten (W) is considered to be the primary solid material choice for divertor components in future fusion reactors due to its high melting point, low sputtering coefficient, high thermal conductivity, low transmutation probability, low tritium retention and good mechanical strength. [1-4] In a fusion reactor, tungsten in the first wall will be directly exposed to 14.1 MeV neutrons, produced by the deuterium-tritium fusion reaction. Collisions between 14.1 MeV neutrons and tungsten atoms produce tungsten primary-knock-on atoms (PKA) with various recoil energies. The PKAs ultimately lose their acquired energy, mostly by displacing other atoms, creating vacancy and self-interstitial atom (SIA) point defect clusters, in a sequence known as a displacement cascade. The continuous creation of point defects and clusters of these defects, their subsequent migration, additional clustering and dissolution of clusters, interaction with preexisting point defects (alloying elements) and extended defects, along with the production of transmutants, results in continual and often dramatic microstructural changes during neutron irradiation. [4] The physical and mechanical properties of materials directly depend on their microstructure. Therefore, the microstructure in tungsten generated due to neutron damage and its effect on the mechanical and thermal properties is a significant issue for fusion power system applications.

---

[1] Corresponding Author  Email address: giridhar.nandipati@pnnl.gov (Giridhar Nandipati)
Tel.: +1(509) 375-2795, fax: +1(509) 375-3033

The object kinetic Monte Carlo (OKMC) method is an excellent method to study the kinetics of competing processes and to isolate their effects on microstructural evolution. As such, the OKMC simulation code, KSOME, described in Refs. [5, 6] was used to carry out the present simulations. The OKMC method has been previously used to study irradiation damage in Fe and Cu [7-10] However, no such studies of radiation damage in tungsten have been done to date. Moreover, to the best of our knowledge, no OKMC simulation studies have been done in any fusion relevant materials to understand the differences in the damage accumulation due to differences in the PKA spectra of radiation sources. This article is Part-I of a 4 Part series of articles on the simulation study of damage accumulation in pure, polycrystalline tungsten during neutron radiation with PKA spectra of the High Flux Isotope Reactor (HFIR) and 14 MeV neutrons, using the OKMC method. Our interest in modeling HFIR irradiation conditions stems from the fact that many experimental studies of radiation effects in tungsten are performed in fission reactors, which differ markedly from the fusion nuclear environment. The objective of this study is to understand the effects of PKA spectra (for both 14 MeV and HFIR neutrons), as well as to understand the impact of intragranular defects and transmutation products on the microstructural evolution of tungsten during neutron irradiation. We present here the simulation results of radiation damage in tungsten due to neutron bombardment with the PKA spectrum of HFIR at various dose rates and grain sizes. To isolate the effects of PKA spectra, in Part-I and Part-II the intragranular traps or pre-existing defects and transmutation are ignored. In Part-II, the simulation results on the radiation damage in tungsten due to bombardment by 14 MeV neutrons are compared to the damage due to neutron radiation with the HFIR PKA spectrum. In Part-III the effects of intragranular traps on the damage accumulation are examined for both PKA spectra, and the influence of transmutation is examined in Part-IV. The only type of extended defect in the present simulation is a grain boundary.

## 2. Simulation Details

In this section, we present in detail the choice of shape and dimensions of the simulation cell, boundary conditions, cascade database and parameters related to the PKA spectrum of HFIR that are used to calculate dose rates and their corresponding cascade production or insertion rates. In the present simulations, the single-vacancy and SIA cluster migration barrier are taken as 1.3 eV and 0.05 eV [11], respectively. The values of the defect binding energies were taken from the *ab initio* calculations of Becquart *et al*. [12] and references therein. Detailed descriptions of all other kinetic parameters used in the present simulations are provided in Ref [5].

*2.1 Simulation Cell*

Simulations were performed using a non-cubic box with dimensions 95.10 x 96.37 x 97.00 nm$^3$ (300$a$ x 304$a$ x 306$a$, where $a$ is the lattice constant of tungsten), with each axis parallel to a <100> direction of the crystal. Each defect is allowed to hop to one of eight body-centered cubic (bcc) nearest neighbor lattice sites along <111> directions. The effect of finite grain size was introduced by adopting finite periodic boundary (FBC) conditions in all three directions.[2] [13, 14] A defect is removed from the simulation when it satisfies two conditions; (1) it crosses a boundary and (2) it has already diffused further than half the grain size (the simulation box is assumed to be at the center of a grain). The selection of grain sizes of 2.0 and 4.0 μm in diameter is based on the Japanese ITER-grade tungsten, which consists of elongated grains that are 5.0 μm in length and 1.0-3.0 μm wide. [15]

Since the simulation box size is much smaller than the grain size, a fast 1D-diffusing SIA cluster can repeat its path and explore the same local landscape repeatedly before the next cascade is inserted. For example, in the case of a cubic box, an SIA cluster repeats its path every two boundary crossings (Fig. 1(d)) and only once if it diffuses along the diagonal of the simulation box. When SIA clusters repeat their path, they tend not to interact with existing defects before they reach a grain boundary, resulting in an over-estimation of the number densities of vacancies and vacancy clusters. Non-cubic boxes (Fig. 1(a-c)) allow 1D-diffusing SIA clusters to explore more space of the simulation box [16] compared to a cubic one. For a non-cubic simulation cell, the number of boundary crossings depends on its dimensions. For example, in the case of a simulation box with dimensions of 300$a$ x 350$a$ x 400$a$, an SIA cluster will have crossing points on the face of the boundary that are separated by a distance of 50 lattice units corresponding to the greatest common factor (GCF) of the box dimensions. Hence, an SIA cluster will cross the shortest cell boundary 6 times and travels a distance of 988nm before it repeats its path. While in a simulation box with GCF = 2 (dimensions of 300$a$ x 302$a$ x 306$a$), an SIA cluster repeats its path after 150 boundary crossings on the shortest dimension of the cell. Note that a simulation box with dimensions as e.g. 300$a$ x 301$a$ x 302$a$ has the ideal GCF=1. However, the "grid method" or "cell-list" algorithm [17] to speed up the search for nearby defects cannot be applied to a box with GCF = 1. Therefore, we chose the simulation box dimensions to be 300$a$ x 304$a$ x 306$a$ with a GCF = 2.

---

[2] In FBC, normal periodic boundary conditions (PBC) are applied when a defect crosses a boundary, but whenever the distance between the starting location of a defect (at the moment of its creation) and its present location is greater than or equal to the radius of the grain size, the defect is removed from the simulation.

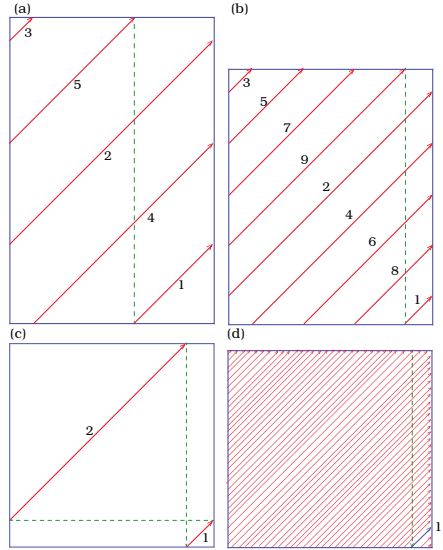

**Figure 1.  Illustration of 1D diffusion in a 2D rectangular box with periodic boundary conditions (a) 20 cm x 30 cm (b) 25 cm x 30 cm (c) 20 cm x 20 cm (d) 20 cm x 19.8 cm (cm = centimeters)**

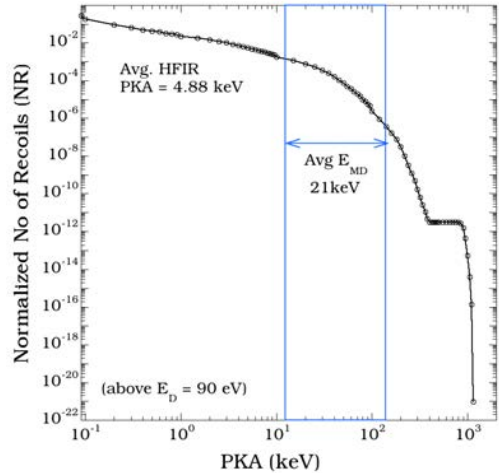

**Figure 2.  PKA spectrum for tungsten under HFIR irradiation conditions**

*2.2 Cascade Database*

A portion of the recoil energy ($E_{PKA}$) is dissipated in electronic losses and the remainder in elastic collisions ($E_{MD}$). The damage energy ($E_{MD}$) is equivalent to the cascade energy in a molecular dynamics (MD) simulation.

$E_{MD}$ for a given $E_{PKA}$ was computed using the SRIM software [18] using the displacement threshold energy, $E_D$ = 90 eV [19], and each calculation is averaged over 5000 recoil simulations. Following Ref. [20], the lattice binding energy was set to zero so that the $E_{MD}$ is the difference between the initial $E_{PKA}$ and the energy dissipated in ionization.

An extensive database of cascades for $E_{MD}$ ($E_{PKA}$) ranging from 10-100 keV (12.5-137.3 keV) was generated using MD simulations. [21] A new database of cascades was created such that the concentration of cascades of a particular $E_{MD}$ is according to the normalized fraction of recoils at the corresponding $E_{PKA}$ in the HFIR PKA spectrum shown in Fig. 2. Due to the small number of Frenkel pairs produced in cascades with $E_{MD}$ lower than 10 keV, they are not included in the database. Defect cluster statistics in the cascade database is tabulated in the Supplemental Information in Ref. [21].

*2.3 Calculation of Dose and Dose Rate*

In OKMC simulations, each point defect cluster is treated as one object characterized by defect's type, size and the position of its center of mass. A cascade, which is a collection of objects (defect clusters) along with their spatial distribution[3], is selected randomly from the database and inserted into the simulation cell at random positions at a rate corresponding to a particular dose rate. The production (insertion) rate of cascades and the accumulated damage, measured as *DPA* (displacements per atom), are calculated based on the NRT displacements per cascade ($\nu_{NRT}$), [22]

$$\nu_{NRT} = 0.8 \frac{E_{MD}}{2E_D} \quad (1)$$

where $E_D$ = 128 eV for tungsten calculated using the EAM potential employed here. [23] It is well known that in MD simulations, the number of created Frenkel pairs[4] depends sensitively on the interatomic potentials. [24, 25] A recent study [26] showed that the defect production curves among bcc metals collapse onto a single curve across a wide range of cascade energies if the cascade energies are scaled with the $E_D$ of the corresponding potential. Therefore, to obtain a more consistent one-to-one comparison between MD's DPA and experimental DPA, the MD defect production should be analyzed with $E_D$ = 128 eV of the potential, while the experimental defect production should be calculated with $E_D$ = 90 eV (typically used to estimate the experimental dose using the SRIM software). Therefore, we use the $E_D$ = 128 eV of the potential to calculate the damage (dpa) in our simulations. Using Eq. 1 and $N_{at}$, which is the number of atoms in the simulation box, DPA per cascade (*DPC*) is calculated as

$$DPC = \frac{\nu_{NRT}}{N_{at}} \quad (2)$$

Note that for a particular dose rate, the cascade production rate depends on $N_{at}$. The $\nu_{NRT}$ and *DPC* calculated using an $E_D$ = 128 eV and $N_{at}$ = 55.814 x 10$^6$ for various $E_{MD}$ energies as shown in Table 1.

**Table 1. NRT displacements per cascade ($\nu_{NRT}$) and DPA per cascade (DPC) in tungsten at various PKA energies from MD simulations**

| $E_{MD}$ (keV) | 10 | 20 | 30 | 40 | 50 | 60 | 75 | 100 |
|---|---|---|---|---|---|---|---|---|
| $\nu_{NRT}$ | 31.25 | 62.50 | 93.75 | 125.0 | 156.25 | 187.50 | 234.38 | 312.50 |
| *DPC* /(10$^{-6}$) | 0.559 | 1.12 | 1.68 | 2.24 | 2.80 | 3.36 | 4.20 | 5.60 |

The average dose rate in HFIR is approximately 10$^{-7}$ dpa/s or 4.73 dpa/year. [27] As the *DPA* is not an observable in experiments, we studied damage accumulation over a range of dose rates. The average $E_{PKA}$ for the HFIR PKA spectrum (see Fig. 2) within the range of 12.5-137.3 keV is approximately 28 keV and the corresponding $E_{MD}$ = 21 keV while the average $E_{PKA}$ for the entire HFIR spectrum is 4.88 keV. For the dose rates of 1.7 x $10^{-5}$ to 1.7 x $10^{-9}$ dpa/s, the corresponding cascade production rates are 0.851 x $10^2$ to 0.851 x $10^{-2}$ cascades per second.

---
[3] Relative positions of defect clusters within the cascade volume, as obtained from MD simulations.
[4] A single, stable interstitial atom and its related vacancy.

## 3. Results and Discussion

### 3.1 General Details

The results presented here are specifically for the defect accumulation in tungsten with grain sizes of 2.0 and 4.0 μm at 1025 K, which is above the Stage III annealing temperature so that single-vacancies are mobile. From MD simulations, the defect cluster sizes and their spatial distribution (morphology) can change significantly over a temperature range from 300 to 2050 K even though the average number of Frenkel pairs does not change significantly. Therefore, the cascades used in the present simulations were specifically created in MD simulations performed at 1025 K. As there are no intragranular traps in our simulation, SIA clusters of all sizes either recombine with vacancy clusters or are absorbed at grain boundaries. Consequently, only the accumulation of immobile vacancy clusters was observed. For all the dose rates studied, the density of vacancies and vacancy clusters decreased with increasing grain size and can be explained based on the fact that the distance an SIA cluster has to travel increases with increasing grain size, thereby increasing the probability of recombination. For all grain sizes studied, the vacancy cluster density increases, the number density of vacancies and, consequently, the average vacancy cluster size decreases with increasing dose rate.

### 3.2 Microstructure of neutron-irradiated pure tungsten at 1.0 dpa

Figs. 3 and 4 show the void microstructures in the 2.0 and 4.0 μm grains for the dose rates of 1.7 x $10^{-5}$ to 1.7 x $10^{-9}$ dpa/s, when viewed from different directions, at 1.0 dpa. The general characteristics of the microstructure are similar for both grains, i.e. that the vacancy cluster size increases while its density decreases with decreasing dose rate, and that the spatial ordering of voids along (011) planes becomes clearer at lower dose rates. Note that the ordering appears along the (011) plane first before it appears along any other {110} plane because of the non-equivalent box dimensions used in this study. Otherwise, the appearance of completely formed planes of vacancy clusters along any of the 6 {110} planes should be equally probable. A closer look at the [010] and [001] projections reveals regions of partial clearing along the (110) planes, indicating that spatial ordering does occur on (110) planes even though it is less visible than the ordering on the (011) planes. The average interplanar spacings along the [011] direction are given in Table.2. The spacing increases with the grain size. However, no particular dependence on dose rate was observed. In addition, the spacing appears more uniform at the lower dose rate and larger grain size (see Fig. 3 and 4). When the grain size is increased to 10.0 μm, the spacing increases even further (not shown).

**Table 2. Inter-planar spacing (in nm) of the void lattice along the [100] direction.**

| Dose Rate (dpa/s) | $10^{-5}$ | $10^{-6}$ | $10^{-7}$ | $10^{-8}$ | $10^{-9}$ |
|---|---|---|---|---|---|
| 2.0 μm | 13.67 | 17.09 | 13.67 | 11.39 | 13.67 |
| 4.0 μm | 22.79 | 17.09 | 17.09 | 22.79 | 22.79 |

Void lattice formation along {110} planes with a mean void size of 4.7 nm and void spacing of 15 nm was observed in neutron-irradiated tungsten at a dose of 1.54 dpa in the JOYO fast test reactor at 750 $^0$C (1023 K). [28-30] Also, a void lattice with lattice parameter of 19.5 nm and mean size of 3.0 nm in tungsten irradiated in EBR-II reactor at 550 $^0$C (823 K) at a dose of ~7.0 dpa was observed. [31] Although, both JOYO (≈$10^{-7}$ dpa/s) [32] and HFIR have similar dose rates, the production of solid transmutants (Re and Os) is approximately 10 times faster in HFIR due to the higher thermal neutron content in its neutron spectrum. [33,34] Production rates of Re are 0.95%/dpa and 9.2%/dpa in JOYO and HFIR, respectively. [35] Re and Os readily form mixed dumbbells [36] which perform 3D diffusion due to low rotation barriers, suppressing the formation of the void lattice. [33] On the other hand, the production of gaseous transmutants (He and H) in HFIR is very small for the doses simulated in this study. [37, 38]

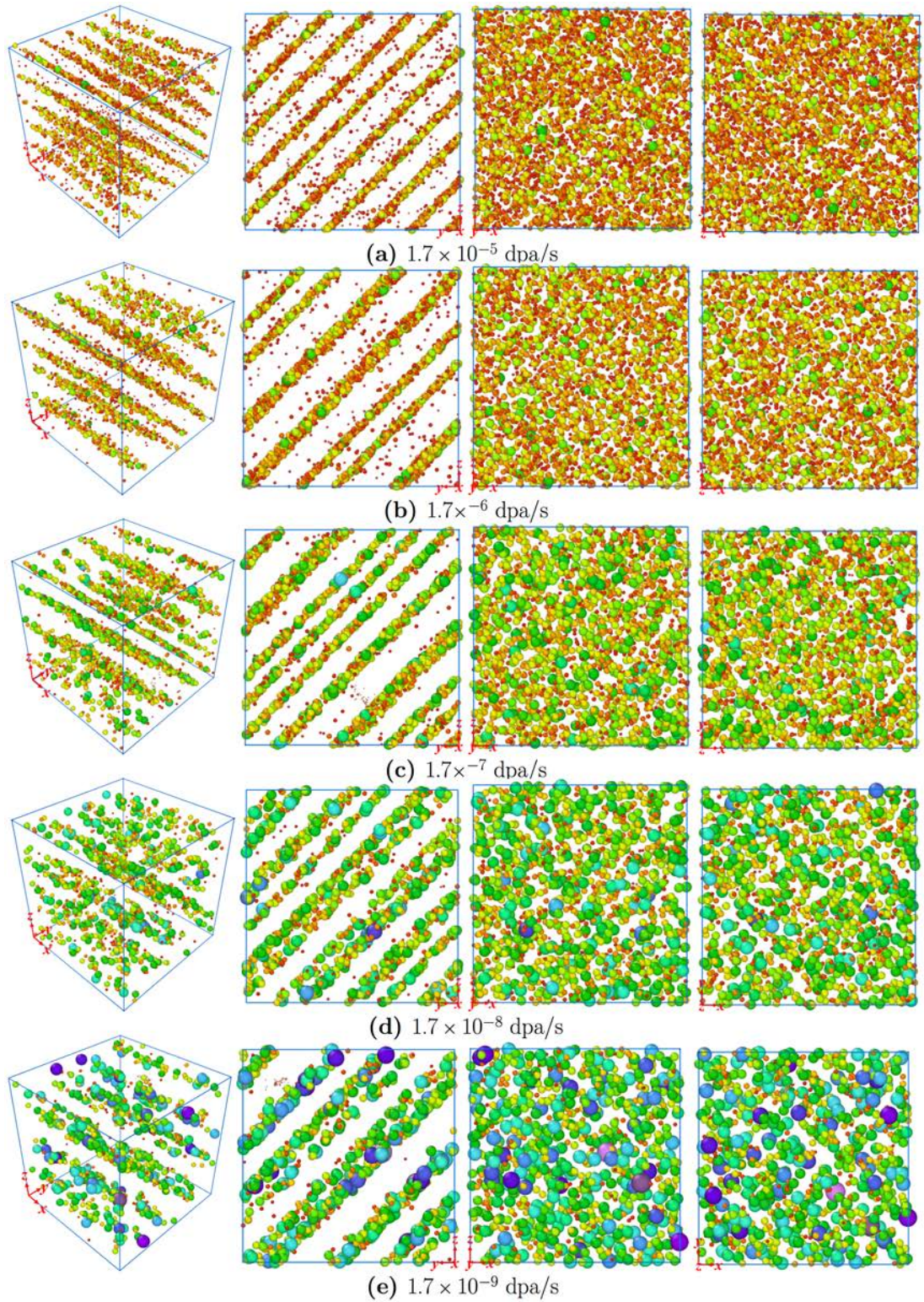

**Figure 3. Microstructure of vacancy clusters at a dose of 1.0 dpa for 2.0 μm grain size when viewed in perspective, and along the [100], [010] and [001] directions at dose rates of (a) $1.7 \times 10^{-5}$ dpa/s (b) $1.7 \times 10^{-6}$ dpa/s (c) $1.7 \times 10^{-7}$ dpa/s (d) $1.7 \times 10^{-8}$ dpa/s (e) $1.7 \times 10^{-9}$ dpa/s.**

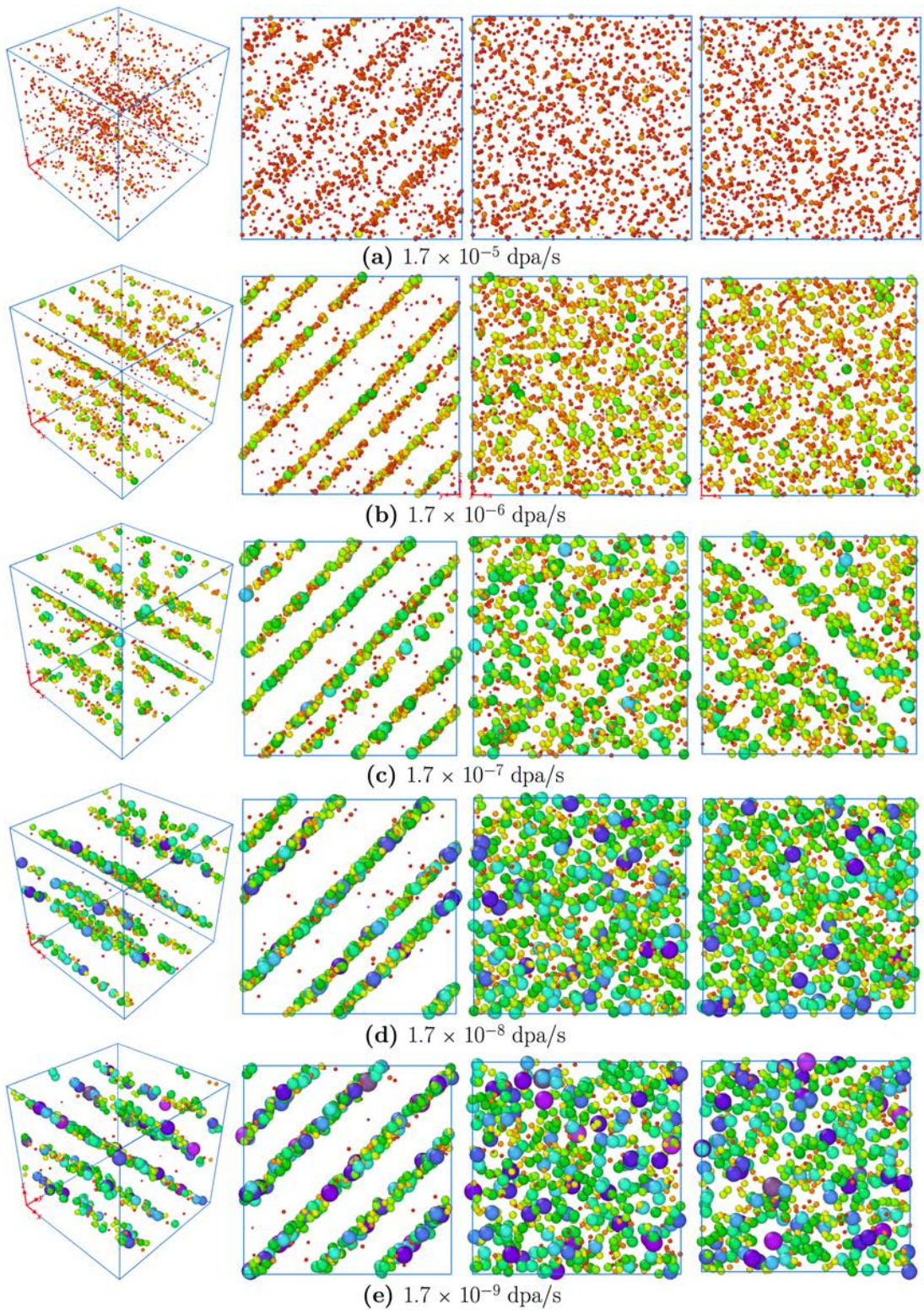

**Figure 4 Microstructure of vacancy clusters at a dose of 1.0 dpa for 4.0 μm grain size when viewed perspective, and along the [100], [010] and [001] at dose rate of (a) $1.7 \times 10^{-5}$ dpa/s (b) $1.7 \times 10^{-6}$ dpa/s (c) $1.7 \times 10^{-7}$ dpa/s (d) $1.7 \times 10^{-8}$ dpa/s (e) $1.7 \times 10^{-9}$ dpa/s.**

It is already well-known that void lattice formation is due to 1D diffusion of SIA clusters. [39-47] However, 1D SIA diffusion alone, while necessary, is not sufficient to form a void lattice. To illustrate, at 1025 K, after 1.0 ns of annealing, the average fraction of SIA clusters (sizes > 5) that diffuse only in 1D, range from as low as 20% for 20 KeV cascades to 70% for 100 keV cascades (see Fig. 9 in Ref. [48]). Yet, a void lattice forms. On the other hand, in a separate set of simulations at 300 K, which is well below Stage III, where all SIA clusters diffuse exclusively in 1D, no spatial ordering of voids was observed. [49] Therefore, void lattice formation also requires the diffusion of mono-vacancies and the dissociation of small vacancy clusters, both of which are active at 1025 K but not at 300 K. Also, in tungsten, the dissociation of small vacancy clusters and inhibition of nucleation of new vacancy clusters due to the negative di-vacancy binding energy appears to hasten the formation of the spatial ordering along {110} planes. More importantly, all surviving vacancy clusters were produced during a cascade event, and none were created due to nucleation.

*3.3 Damage Accumulation*

Figs. 5(a-h) show the density of vacancies and vacancy clusters, the average vacancy cluster diameter, and the fraction of vacancies in visible clusters, as a function of dose and dose rate for 2.0 μm and 4.0 μm grains. Spherical clusters are assumed in calculating their diameter. The most significant effect of an increase in the grain size is a decrease in the densities of both single vacancies and vacancy clusters. The qualitative behavior of defect cluster evolution in both grains as a function of dose and dose rate remains the same.

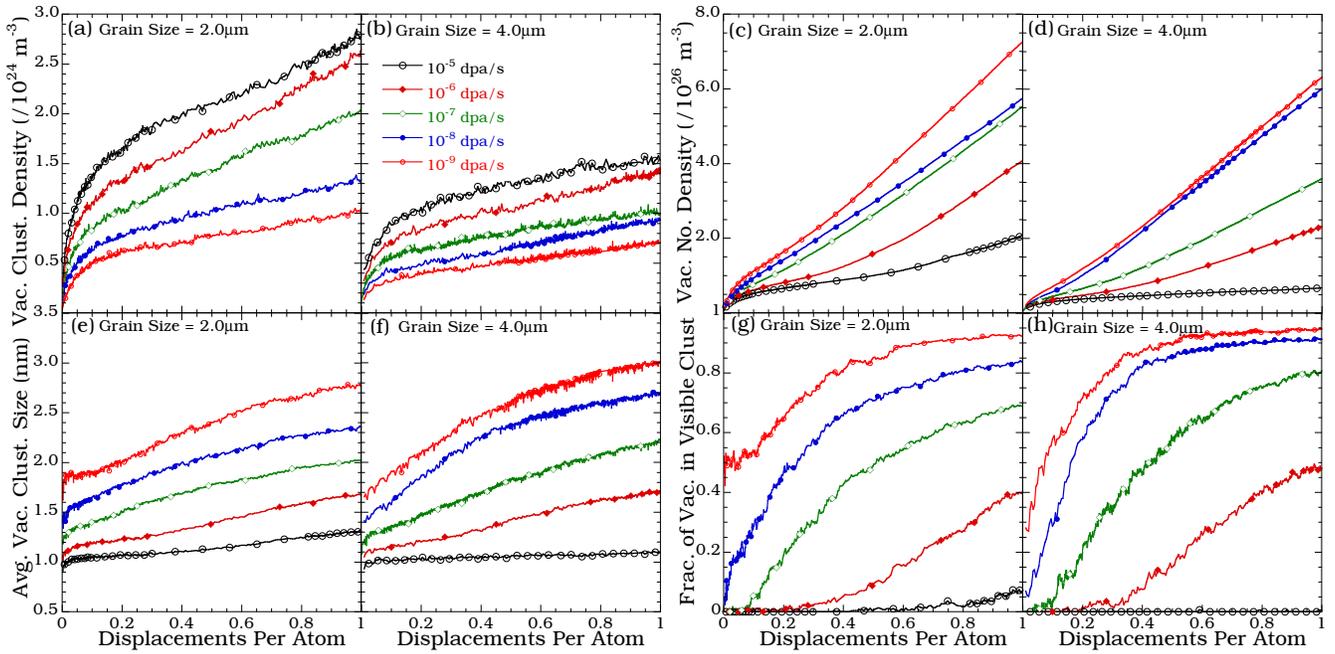

**Figure 5.** (a-b) Density of vacancies; (c-d) Density of Vacancy Clusters; (e-f) Average vacancy cluster size (diameter); (g-h) Fraction of vacancies in visible clusters (sizes larger than 2 nm), as a function of dose for various dose rates.[5]

Another intriguing result is that, with increasing grain size, there is a slight increase in the fraction of vacancies in visible clusters and the average vacancy cluster size for all dose rates studied. An exception is at the highest dose rate of 1.7 x $10^{-5}$ dpa/s, in which the opposite trend is observed. This behavior is accompanied by a sharper cluster size distribution at higher dose rate (more homogeneous), which becomes wider with decreasing dose rate (Fig. 6) (Ostwald ripening). For the dose rate of 1.7 x $10^{-5}$ dpa/s the average vacancy cluster size remains constant with dose for the grain size of 4.0 μm (see Fig. 5(f)).

For a given dose rate, how the Ostwald ripening appears to increase in larger grains (pseudo-ripening) is not immediately clear. In fact, the process is controlled by the SIA clusters. In a larger grain, the SIA clusters effectively live longer before they are absorbed by the grain boundary. Indeed, as will be described in more

---
[5] For brevity the factor of 1.7 is omitted in Table. 1 and, Figs 5 and 6. Therefore, the value of $10^{-5}$ dpa/s represents a dose rate of $1.7 \times 10^{-5}$ dpa/s.

detail below, it is observed that as the vacancy cluster size distribution becomes wider, the probability of recombination of small vacancy clusters with SIA clusters increases. Therefore, having SIA clusters live longer in larger grains, the contribution of the pseudo-ripening to the overall ripening increases. Experimentally, a decrease in the density of vacancy clusters with increasing grain size was observed in Cu irradiated with fission neutrons in DR-3 reactor at 623 K up to 0.3 dpa [50]

During the initial stages of irradiation, damaged regions have large spatial separation resulting in a significant fraction of SIA clusters diffusing to grain boundaries, thereby making a minimal contribution towards intercascade recombination. This results in a sharp increase in the density of vacancies and vacancy clusters (damage accumulation). With continued irradiation, the spatial separation between vacancy clusters decreases, resulting in an increase in the fraction of SIA clusters contributing to intercascade recombination leading to a slow down in the rate of damage accumulation. From Figs. 5(a-d) one can see the slowdown in the accumulation rate after an initial sharp increase indicated by the first inflection point at lower doses. If the spatial distribution of vacancy clusters is random, with increasing dose almost all SIA clusters recombine before they reach a grain boundary, resulting in saturation of the density of vacancies and vacancy clusters. However, in the present simulations, due to the spatial ordering of vacancy clusters, their density starts to increase again at a higher dose after the initial slow down. The spatial ordering of vacancy clusters results in the creation of regions with high and low vacancy cluster densities as well as vacancy-free regions, which results in the decrease of intercascade recombination and an increase in the fraction of SIA clusters escaping to grain boundaries when compared to vacancy clusters distributed randomly in space. In the case of 4.0 μm grain size and the dose rate of $1.7 \times 10^{-5}$ dpa/s (see Figs. 4(a) and 5(b, d, f)), the average vacancy cluster size remains constant, while the densities of vacancies and vacancy clusters increase very slowly (appearing almost constant), until the distinct appearance of the void lattice. This shows that the appearance of the void lattice is pushed to higher doses with increasing grain size and dose rate.

Figs. 5(a-d) show that the density of vacancies decreases while the density of vacancy clusters increases with dose rate for both grain sizes. Consequently, the average cluster size decreases with increasing dose rate (see Figs. 5(e-f)). In addition, there is a reduction in the growth rate of the density of vacancies with increasing dose rate (see Figs. 5 (c-d)). Figs. 5(g-h) show an increase in the fraction of vacancies that are part of visible clusters[6] and the onset of the appearance of visible clusters also shifts to higher doses with increasing dose rate. Also, the fraction of vacancies in visible clusters eventually saturates, and the onset of the saturation moves to a higher dose with both increasing dose rate and grain size. Accordingly, it can be inferred that the fraction of vacancies in the invisible clusters should also saturate. Similar to average vacancy cluster size, the fraction of vacancies in visible clusters increases with grain size at lower dose rates ($\leq 1.7 \times 10^{-7}$ dpa/s) and decreases for higher dose rates ($1.7 \times 10^{-4}$ dpa/s). Void density and the average void diameter in a recent experimental study of pure neutron-irradiated tungsten at 1073 K in HFIR up to 0.98 dpa are $0.8 \times 10^{22}$ m$^{-3}$ and 3.8 nm, respectively [35,36]. In contrast, void densities in our simulations are approximately an order of magnitude higher. It is possible that this could be due to a larger average grain size of the tungsten sample used in the experiments and also due to transmutation which is shown to suppress the void formation. [51,52]

At 1025 K, both the diffusion of mono-vacancies and dissociation of small vacancy clusters are active, and the coarsening of vacancy clusters becomes the dominant process. SIAs are either recombined or absorbed at the grain boundary following a cascade insertion. Moreover, at lower dose rates, the time-interval between cascade insertions is large enough that even mono-vacancies, either produced during a cascade event or from the dissociation of small clusters, have sufficient time to diffuse the micron distances required to reach a grain boundary during the initial-stage of irradiation (at very low dose). However, with increasing dose, mono-vacancies are captured by larger vacancy clusters, resulting in the growth of large vacancy clusters at the expense of small clusters, consistent with Ostwald ripening. [53,54] With decreasing dose rate, vacancy clusters have a longer time interval to coarsen between cascade insertions, resulting in faster growth of large vacancy clusters (see Fig. 5(e-f)); or more simply stated, the vacancy cluster sink strength for vacancy absorption is much stronger than the grain boundary sink strength. Vacancy cluster coarsening between cascade insertions reduces the total number of defect clusters, and, in turn, the number of recombination events. As such, the density of vacancies increases with decreasing dose rate (see Fig. 5(a-b)). Altogether, both the vacancy cluster size and the density of vacancies increase, while the density of vacancy clusters decreases with decreasing dose rate (*See* Fig. 5(a-f)).

---

[6] A vacancy cluster with a diameter larger than 2.0 nm (approx. 300 vacancies) is considered visible under TEM (Transmission Electron Microscope) examination.

Fig. 6(a-b) shows the vacancy cluster size distribution as a fraction of vacancy cluster density, with a bin size of 100 vacancies or a diameter of 1.45 nm, at a dose of 1.0 dpa for various dose rates at grain sizes of 2.0 and 4.0 μm. It clearly shows that the vacancy cluster size distribution becomes wider with decreasing dose rate and increasing grain size. At $1.7 \times 10^{-5}$ dpa/s, a large fraction of vacancy clusters are smaller than 300 vacancies or a diameter of 2.0 nm, but with decreasing dose rate the vacancy cluster size distribution becomes wider. However, for both grain sizes, the fraction of small vacancy clusters which are smaller than the average cluster size for all dose rates is significant. Therefore, as the distribution becomes wider, the probability of recombination of small vacancy clusters with an SIA cluster increases. With increasing grain size, this results in an increase in the average cluster size and the fraction of vacancies that are part of visible clusters at lower doses rates, while for higher dose rates they decrease.

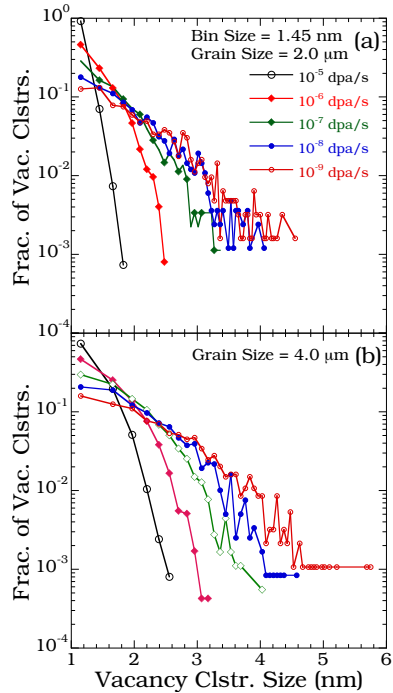

**Figure 6. Binned vacancy cluster size distribution at a dose of 1.0 dpa, for grain sizes of (a) 2.0 μm (b) 4.0 μm**

*3.3 Effect of Temperature*

To understand the effect of temperature on defect accumulation, ad hoc simulations were performed at 300 and 2050 K using cascades of $E_{MD}$ equal to 75 and 100 keV and a wider range of dose rates than those studied here ($10^{-2}$ to $10^{-8}$ dpa/s). At 2050 K, no defect accumulation was observed, as both mono-vacancy diffusion and dissociation of vacancy clusters are very active. More importantly, cascades produced at 2050 K have minimal in-cascade vacancy clustering [21], and the negative binding energy between mono-vacancies prevents nucleation of any new vacancy clusters. [12] Consequently, all vacancy clusters quickly dissociate into mono-vacancies, which readily diffuse out to grain boundaries long before the next cascade is inserted. Note that damage evolution is much faster than the cascade insertion rate as the higher temperature drives sufficient defect kinetics to rapidly approach thermodynamic equilibrium. Moreover, to see any effect, it is expected that dose rates have to be 1.0 dpa/s or higher. This behavior would be true even for most reactor conditions, unless impurities stabilize di-vacancies from dissociating, which would be necessary to nucleate larger clusters for defect accumulation.

At 300 K, neither the diffusion nor the dissociation of vacancies is active. Consequently, all vacancy clusters are randomly distributed in the simulation cell, and a large fraction of them exist as mono-vacancies. [49] The average vacancy cluster size shifts to values closer to a mono-vacancy with decreasing $E_{MD}$. Moreover, within the range studied, dose rates appear to have no effect on vacancy defect accumulation or their size distribution. Consistent with the review of experimental results presented in Refs. 39 and 40, the KMC simulations at 300 K do not indicate void lattice formation. Similar behavior is expected for both HFIR and 14 MeV PKA spectra for the range of dose rates studied in the present simulations.

At both 300 K and 2050 K, damage evolution ceases as soon as SIA clusters and all mobile vacancies are absorbed in grain boundaries, respectively. The absence of the effect of dose rate is because the time-interval between cascade insertions is longer than that for the damage evolution, i.e., at both 300 K and 2050 K, within the time-scale of a simulation (or the maximum length of simulated time), is faster than the cascade insertion rate. Within the time-scale of a simulation, to observe any effect of dose rate on the damage accumulation, the cascade insertion rate has to be at least equal to the slowest diffusing defect species.

**4. Conclusions**

OKMC simulations, using KSOME, were carried out to examine damage accumulations in polycrystalline bulk tungsten irradiated at various dose rates and grain sizes according to the PKA spectrum corresponding to HFIR, at 1025 K. Both dose rate and grain size have significant effects on the accumulation of damage. For a given dose, increasing the dose rate has the following consequences: 1) the density of vacancies decreases, 2) the average size of vacancy clusters decreases, and 3) the density of vacancy clusters increases. The growth of large vacancy clusters increases with decreasing dose rate and is mainly due to the coarsening or Ostwald ripening of vacancy clusters in the time interval between cascade insertions.

A pseudo-ripening effect that is caused by the longer effective lifetime of SIA clusters was observed in larger grains before they were absorbed at the grain boundaries. Consequently, the increasing probability of recombination with vacancy clusters (particularly the small vacancy clusters) leads to a pseudo-ripening effect. Ostwald ripening and pseudo-ripening result in a decrease of the density of vacancies and vacancy clusters with increasing grain size.

The spatial ordering of vacancy clusters along {110} planes was observed at all the dose rates studied, but their appearance shifts to a higher dose with increasing grain size. The interplanar spacing appears to be random with no particular dependence on the dose rate, but it increases with grain size. In relation to the void lattice formation mechanism, we observed the following important findings:

1) While 1D diffusion of SIA clusters is necessary, it is not sufficient. Additionally, diffusion of mono-vacancies is required to form a void lattice, and the dissociation of small vacancy clusters will accelerate the appearance of ordering.
2) At 300 K, despite the fact that all SIA clusters diffuse 1D, a void lattice does not form because mono-vacancies cannot diffuse.
3) At 2050 K, no accumulation of damage will be observed (hence no void lattice) because the primary defect state of the cascade does not contain large enough vacancy clusters to stabilize inside the grain before the defects migrate to the grain boundaries.

4) We describe how to choose the simulation box dimensions to carry out OKMC simulations correctly with 1D-diffusing of SIA clusters.

**ACKNOWLEDGEMENT**

The work described in this article was performed at Pacific Northwest National Laboratory, which is operated by Battelle for the United States Department of Energy under Contract DE-AC06-76RL0-1830. The U.S. Department of Energy, Office of Fusion Energy Sciences (FES) and Office of Advanced Scientific Computing Research (ASCR) has supported this study through the SciDAC-3 program. All computations were performed on CARVER and HOPPER at National Energy Research Scientific Computing Center (NERSC) and Argonne National Laboratory's MCS Workstations. The authors would like to thank Larry Greenwood of PNNL for providing the PKA spectrum for HFIR. Also, the authors would like to acknowledge the use of OVITO [55] for visualization.